\pdfoutput=1
\documentclass[11pt]{article}

\usepackage[final]{acl}

\usepackage{times}
\usepackage{booktabs}
\usepackage{latexsym}

\usepackage[T1]{fontenc}
\usepackage[utf8]{inputenc}

\usepackage{microtype}

\usepackage{inconsolata}

\usepackage{graphicx}
\usepackage{listings}
\usepackage{xcolor}

\usepackage{amsmath,amsfonts,bm}

\def\eqref#1{equation~\ref{#1}}
\def\1{\bm{1}}

\DeclareMathAlphabet{\mathsfit}{\encodingdefault}{\sfdefault}{m}{sl}
\SetMathAlphabet{\mathsfit}{bold}{\encodingdefault}{\sfdefault}{bx}{n}

\definecolor{keywordcolor}{rgb}{0,0,0.6}
\definecolor{stringcolor}{rgb}{0.6,0,0}
\definecolor{commentcolor}{rgb}{0,0.5,0}
\lstdefinelanguage{JSON}{
    basicstyle=\ttfamily\small,
    numbers=left,
    numberstyle=\scriptsize\color{gray},
    stepnumber=1,
    numbersep=8pt,
    showstringspaces=false,
    breaklines=true,
    frame=none, % remove border
    backgroundcolor=\color{white}, % transparent background
    literate=
     *{0}{{{\color{blue}0}}}{1}
      {1}{{{\color{blue}1}}}{1}
      {2}{{{\color{blue}2}}}{1}
      {3}{{{\color{blue}3}}}{1}
      {4}{{{\color{blue}4}}}{1}
      {5}{{{\color{blue}5}}}{1}
      {6}{{{\color{blue}6}}}{1}
      {7}{{{\color{blue}7}}}{1}
      {8}{{{\color{blue}8}}}{1}
      {9}{{{\color{blue}9}}}{1}
      {:}{{{\color{black}{:}}}}{1}
      {,}{{{\color{black}{,}}}}{1}
      {\{}{{{\color{black}{\{}}}}{1}
      {\}}{{{\color{black}{\}}}}}{1}
      {[}{{{\color{black}{[}}}}{1}
      {]}{{{\color{black}{]}}}}{1},
}

\lstdefinestyle{plain}{
  basicstyle=\ttfamily\small,
  breaklines=true,
  frame=none, % remove border
  showstringspaces=false,
  numbers=none,
  tabsize=2,
  language={},
  xleftmargin=0pt, % no left indent
  xrightmargin=0pt % no right indent
}

\usepackage{hyperref}
\usepackage{url}
\usepackage{tabularx}
\usepackage{caption}
\usepackage{appendix}
\usepackage{hyperref}
\usepackage{multicol,multirow}
\usepackage{color-edits}
\addauthor{todo}{red}
\addauthor{zw}{orange}
\addauthor{et}{blue}
\addauthor{gn}{magenta}
\usepackage{xspace}
\newcommand{\ds}{\textsc{Fail-TaLMs}\xspace}
\newcommand{\qaq}{\textsc{Aah}\xspace}
\usepackage{listings}
\usepackage{subfig}
\usepackage{amsmath}
\usepackage{booktabs}
\usepackage{wrapfig}
\definecolor{lightgray}{gray}{0.9}

\usepackage{tcolorbox}
\title{Benchmarking Failures in Tool-Augmented Language Models}

\author{Eduardo Treviño\thanks{Co-First Author} \quad {\bf Hugo Contant}$^*$ \quad {\bf James Ngai} \\ {\bf Graham Neubig} \quad {\bf Zora Zhiruo Wang} \\
Carnegie Mellon University \\
\texttt{\{eatrevin,hcontant\}@andrew.cmu.edu}}

\begin{document}
 \maketitle
\begin{abstract} 

The integration of tools has extended the capabilities of language models (LMs) beyond vanilla text generation to versatile scenarios. However, tool-augmented language models (TaLMs) often assume `perfect' information access and tool availability, which may not hold in the real world.
To systematically study TaLMs' imperfections, we introduce the \ds benchmark, featuring two major failures: under-specified user queries and non-available tools. \ds contains 1,749 examples using 906 tools across 21 categories, including single- and multi-tool usage. 
We evaluate top-performing proprietary and open-source models, and find all current models except for Claude struggle to recognize missing tools or information. Further, to study possible mitigation of the failures, we enable real-time human interaction, named the Ask-and-Help (\qaq) method, to provide missing information or replace non-functional tools. While \qaq can help models solve tasks more correctly when queries are under-specified, it brings minimal benefit when complex tools are broken.\footnote{\href{https://github.com/EduardoTrevino/fail-talms}{https://github.com/Fail-TaLMs}}

\end{abstract}

\section{Introduction}

Tools can greatly enhance language models (LMs) by facilitating their problem-solving process \citep{qin2023toollearning, mialon2023augmentedllms} and extending their abilities \citep{wang2024tools}. Given a user query, a tool-augmented language model (TaLM) can selectively call tools to gather more information and perform computation activities to accomplish the user's request. Such TaLMs have been applied in various scenarios, including interacting with versatile knowledge bases \citep{lazaridou2020opendomainqa}, real-world data \citep{xu2023toolmanip}, and even multi-modal information \citep{gupta2022vp, wang2024trove}.
On the other hand, tools can deprecate over time \citep{qin2023toolllm}, suddenly break \citep{guo2024stable}, or even return unpredictably false outputs \citep{sun2024toolsfaildetectingsilent}.

\begin{figure}[t!]
\centering
\includegraphics[width=0.5\textwidth]{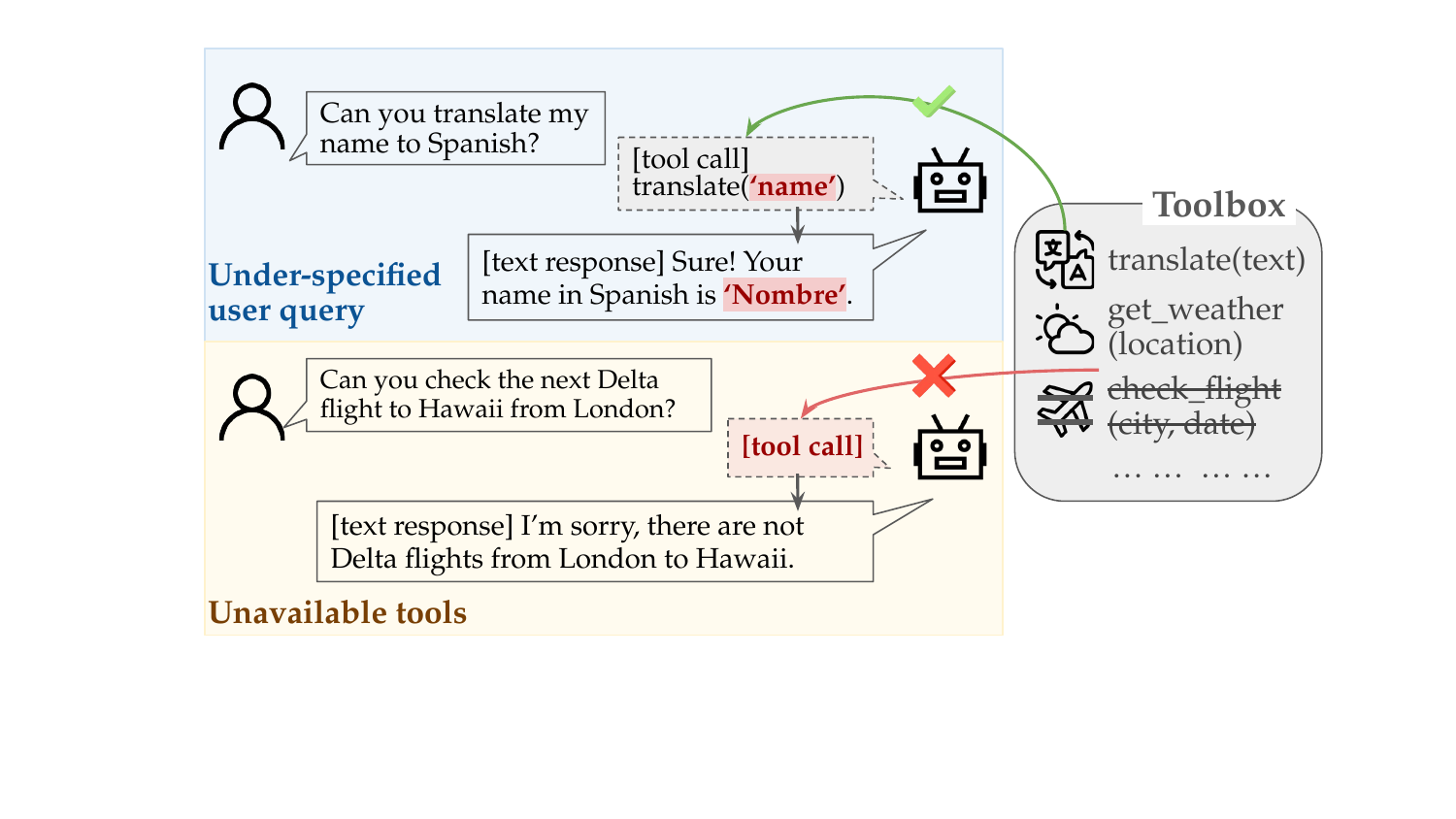}
\caption{Illustration of two major TaLM issues. Left: The user provides \textit{under-specified queries}, which may cause models to hallucinate or make false assumptions about the missing information, e.g., translate the ``name'' string instead of the user's actual name. 
Right: Necessary \textit{tools are unavailable}, e.g., missing \texttt{check\_flight} causes TaLMs to lack the ability to solve the task.}
\label{fig:taLM_example}
\vspace{-4mm}
\end{figure}

Nonetheless, many approaches assume two idealized conditions for TaLM systems: (i) user queries are always sufficiently detailed for models to solve the task, and (ii) all necessary tools are available. In practice, however, these assumptions often do not hold, leading to failures like those depicted in \autoref{fig:taLM_example}. In the first example, the user query is under-specified, and the model, while successfully calling the translation tool, lacks the necessary input, i.e., the user's actual name. As a result, the model incorrectly translates the phrase ``my name'', rather than the intended name. In the second case, the necessary \texttt{check\_flight} tool to solve the task is unavailable, as denoted by the red cross.
In these cases, existing TaLMs often hallucinate contexts or terminate without signaling failure or seeking alternative strategies, leading to sub-optimal behaviors.

To systematically study these practical failures and enable more robust TaLM systems, we introduce \ds~--- a benchmark designed to examine TaLMs under information insufficiency and tool unavailability (\S\ref{sec:3:benchmark}).
We gather 906 real-world tools across 21 categories directly from their host sources, and construct execution environments for all tools along with verifying test cases. Unlike existing benchmarks with limited tool calls via third-party platforms, our tool environment allows real-time and reproducible testing. 
With this collection of tools, we created 575 examples with perfect information and tool availability. We then transformed them into 599 and 575 examples with under-specified queries and unavailable tools, by removing key information from the queries and masking out necessary tools from the provided list, to study the two major failure modes mentioned above.
Overall, \ds contains 1,749 queries across 906 tools from 21 categories.

We propose three evaluation metrics to study model performance: {\it pass rate} to measure task success, {\it awareness} of missing tools or information, and further, {\it unexpected outcomes} to capture when TaLM correctly yet unexpectedly solves a task.

We experiment with a series of top-performing LMs (\S\ref{sec:5:results}): open-weight \textsc{Llama 3} models from 8B, 70B to 405B, and proprietary models including \textsc{Claude} and \textsc{GPT}.
Our experiments reveal that most models struggle to identify the lack of tools or information needed to solve a task, except for \textsc{Claude} with a 56\% awareness rate, 28--54\% higher than other models.
Nonetheless, high awareness does not translate to higher pass rates. For example, GPT-4o achieves 4\% higher pass rate than Claude-3.5-sonnet, despite scoring 44\% lower in awareness.

Finally, to examine whether simple mitigation measures could address these issues, we study if a method enabling TaLMs to interact with humans, dubbed ``Ask-and-Help'' (\qaq), could help obtain missing information or fulfill the function of unavailable tools (\S\ref{sec:4:qaq-method}). 
We measure {\it interaction ratio} to see how often TaLM interacts with humans. 
\qaq substantially improves pass rate particularly when user queries are under-specified, where the models actively interact with humans 21--61\% of the time to gather missing information. However, this human assistance does not bring improvements when tools are unavailable, regardless of the tool functions are replaceable by humans or not, indicating room for better methodologies.

\section{Problem Statement}
\label{sec:2:problem-statement}

A tool-augmented language model (TaLM) consists of (1) a backbone language model $\mathcal{M}$ and (2) a set of tools $T = \{t_1, \dots, t_n\}$. Each tool $t_i$ is a callable function (e.g., \texttt{calculator(expr)}, \texttt{document\_retriever(query,docs)}). Given a natural language (NL) query $q$, the TaLM selects a series of tools $T^q \subseteq T$ to solve the query. For each chosen tool $t_i \in T^q$, the TaLM produces a tool-calling program $p_i^q = \mathcal{M}(q, T)$ that is executed then yielding output $e_i^q$. Finally, the TaLM produces the final answer $r^q$ to query $q$ based on all tool outputs $\{e^q_i\}$.

However, this classic TaLM pipeline may not successfully execute in practice due to two primary issues:

\paragraph{Under-specified Queries}
When the user query $q$ is under-specified, either the subset of relevant tools $T^q$ cannot be successfully identified, i.e., $\mathcal{M}_{\text{det}}(q, T) \not\Rightarrow T^q$, or the set of tool-calling programs $P^q = \{p_i^q\}$ cannot be properly constructed due to insufficient information to determine the input arguments of tools.

\paragraph{Unavailable Tools}
Even when there is sufficient information, the tool may be unavailable due to reasons such as deprecated functions or server execution errors (e.g., server timeout or connection failure). In such cases, the tool execution fails or returns incorrect results, i.e., $\text{exec}(p) \not\rightarrow e$, leading to invalid or inaccurate tool execution outputs.
\section{The \ds Benchmark}
\label{sec:3:benchmark}

In this section, we first introduce the tool collection (\S\ref{sec:3.1:tool-collection}) and benchmark curation processes (\S\ref{sec:3.2:benchmark-creation}), present the data overview (\S\ref{sec:3.3:ds-overview}), then establish the set of evaluation metrics (\S\ref{sec:3.4:eval-metrics}).

\begin{figure*}[t] 
    \centering
    \includegraphics[width=\textwidth]{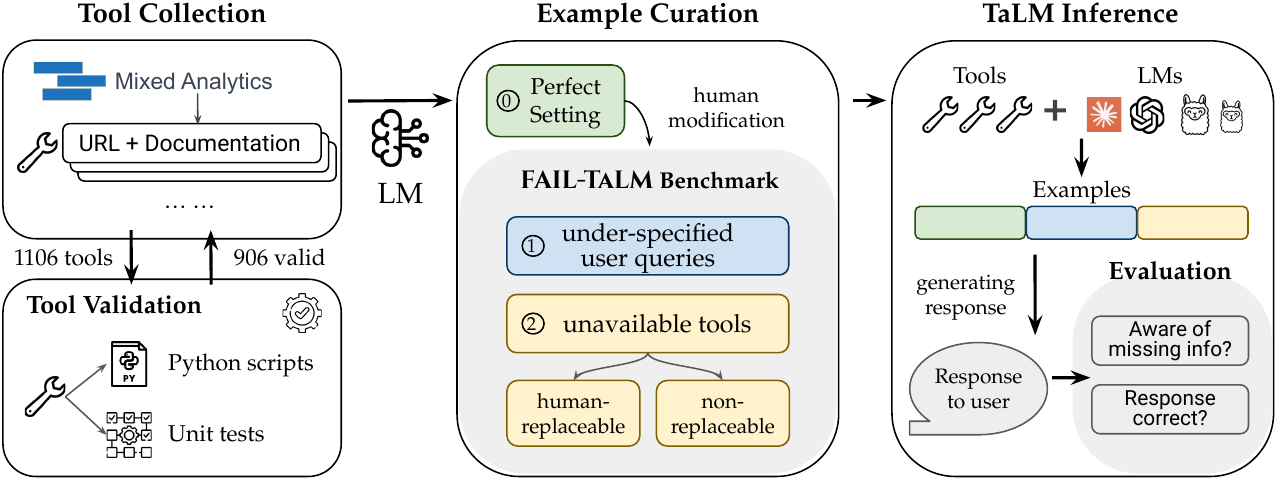}
    \caption{Visualization of the benchmark and tool environment construction (top), as well as the inference pipeline with awareness querying and human interaction phases (bottom).}
    \label{fig:pipeline}
\end{figure*}

%%%%%%%%%%%%%%%%%%%%%%%%%%%%%%%%%%%%%%%%%%%%%%%%%%%%%%%%%%%%
\subsection{Tool Collection and Validation} 
\label{sec:3.1:tool-collection}
% We collect API tools, validate their functions, and construct environments to support their execution. 

\paragraph{Tool Collection}
We use Mixed Analytics\footnote{\url{https://mixedanalytics.com/blog/list-actually-free-open-no-auth-needed-apis/}} and collect 1,106 authorization-free tools, each with an URL, documentation of functionality and argument descriptions, and exemplar use cases.
% As illustrated in the `Tool Collection' block in \autoref{fig:pipeline}

\paragraph{Tool Validation}
To verify the successful executions of collected tools, we transform each tool instance into a callable Python function and synthesize multiple unit tests for it. More specifically, we provide the above-gathered information of each tool to GPT-4o and prompt it (see exact prompt in \S\ref{app:a.1:tool-environment-prompt}) to generate (i) Python file sending requests to the tool URL, and (ii) a JSON schema with the tool's metadata, i.e., tool representation, and (iii) unit tests for each tool. See examples of (i)--(iii) in \S\ref{app:a.2:tool-request}, \S\ref{app:a.3:tool-representation}, and \S\ref{app:a.4:unittest}.

After tool environment construction, we validate if tools can (i) successfully execute and produce valid output and (ii) correctly pass all synthesized unit tests. We only keep tools that encounter no issues in (i) and (ii). Further, we maintain test cases to check tool availability in real-time, with an average response time of $1.47$ seconds. We filter out tests having response times over 20 seconds to enable fast tool responses and test efficiency during inference. After this process, we collected a total of 906 valid tools.

%%%%%%%%%%%%%%%%%%%%%%%%%%%%%%%%%%%%%%%%%%%%%%%%%%%%%%%%%%%%
\subsection{Benchmark Creation}
\label{sec:3.2:benchmark-creation}

Given the tools, we now create queries that ask to solve certain tasks by using one or multiple tools.

\subsubsection{The Standard `Perfect' Setting} 
The standard `perfect' setting adopted by most tool benchmarks assumes fully specified queries and the availability of all necessary tools to solve a given task. We refer to this ideal baseline scenario as the {\it perfect} setting, which serves as the foundation for generating the remaining data settings.

To generate a `perfect' example with query and involved tools, we first construct a set of tool combinations to instantiate NL queries from, by pairing every tool with all other tools within the same category. This systematic approach ensures the usage of every possible tool combinations, instead of biasing over any specific tool. 

% We represent each tool combination with their tool representation and Python unit test. The tool's representation includes (i) required parameters, (ii) optional parameters, (iii) A textual description of the tool's functionality, and (iv) sample responses, see concrete example in \S\ref{app:a.3:tool-representation}. We further add Python unit tests to provide exemplar validated tool usage (\S\ref{app:a.4:unittest}). All information aims to generate tool-using queries that are syntactically correct and executable.

To create an NL query for each given tool combination, we provide the tool information as collected in \S\ref{sec:3.1:tool-collection}, as well as a one-shot (query, tools) example as in \S\ref{app:a.5:perfect-query-example} to demonstrates a query and two tools necessary to solve the query. We instruct GPT-4o to generate queries in realistic usage with content related to the tools' functions, and include the model prompt in \S\ref{app:a.6:perfect-query-prompt}. 

After this step, we perform an additional human validation step. During this step, human reviewers manually examine the generated queries and tool combinations to ensure that the queries are coherent, the arguments provided are valid, and the tool usage is contextually appropriate.
We generate queries for all possible pairs of unique tool combinations in a given category, which yields 575 (query, tools) examples. This serves as the foundation for creating the rest of the benchmark, as illustrated in the \textit{perfect setting} in \autoref{fig:pipeline} the Example Generation module.

\subsubsection{Under-Specified Queries} 
In real-world scenarios, queries are not always fully specified, and crucial details may be omitted. Hence, we create this data split to study whether the model can identify the missing information needed to construct tool calls. We refer to such queries, which maintain their semantic intent but lack essential details, as {\it under-specified queries}.

To create this setting, we modify the \textit{perfect} queries by manually masking out key information required to define the input arguments for the relevant tools. For example, the standard query ``What is the weather in Pittsburgh?'' calls for the tool \texttt{Weather(location: str) $\rightarrow$ str}, which needs the location ``Pittsburgh'' as an argument. By removing ``Pittsburgh,'' the query becomes ``What is the weather?'', which still implies the use of the \texttt{Weather} tool but omits the specific location. We manually remove these critical arguments from the \textit{perfect} queries, yielding a total of 599 under-specified queries. We generate more under-specified queries than perfect ones because a query may have more than one argument and can therefore be masked in multiple ways.

\subsubsection{Unavailable Tools}
In practice, tools required by TaLMs may not always be reliable, for example, tools can be susceptible to depreciation or errors (e.g., 404) especially when provided by third-party platforms with access limitations \citep{guo2024stable}. This data split investigates how models perform when tools turn unexpectedly unavailable. Similarly, we manually decide which tools to remove and verify the quality of the modifications. Moreover, we study the distinction among tools, particularly in whether they can be easily replaced by an average human. We categorize data into two scenarios --- {\it human-replaceable tools} and {\it non-replaceable tools}.

\noindent {\bf Human-replaceable tools} features tools whose functions can be easily replaced by a normal human with minimal effort, such as calculating the value of $1+1$ with a \texttt{calculator} tool, or saying a random joke with the \texttt{joke} tool. Human-replaceable tools often have relatively easy functions. 

\noindent {\bf Non-replaceable tools} usually possess more complex functionalities that humans cannot easily replicate, such as complex calculations (e.g., \texttt{calculator(45465 * 5487)}) or simulating a rocket launch using the \texttt{rocket\_simulator} tool. 

For both unavailable tool scenarios, we manually classify a tool into human-replaceable or non-replaceable by asking 'Can human replace the tool with minimal effort?' If yes, the tool is placed into the {\it{human-replaceable}} setting; otherwise, it is placed in the {\it{non-replaceable}} setting. 
We yield 261 and 314 examples with unavailable tools that are human-replaceable and non-replaceable, respectively. In experiments, we provide all tools except the selected unavailable ones to the TaLM.

\paragraph{No Tools}  \texttt{TaLMs} may know certain unprovided tools, which we are unaware of due to their proprietary training data \citep{zhuang2023toolqa, huang2024meta}, we introduce a \textit{no-tool} setting, where only NL query is provided. This setting serves as a baseline for measuring the model's inherent knowledge. Under this setting we only provide the \textit{perfect} setting queries to TaLM without any tool information. We run the same number of 575 queries as in the \textit{perfect} set.

%%%%%%%%%%%%%%%%%%%%%%%%%%%%%%%%%%%%%%%%%%%%%%%%%%%%%%%%%%%%
\subsection{Benchmark Analysis}
\label{sec:3.3:ds-overview}
As shown in \autoref{fig:data-dist}, our \ds benchmark spans 21 categories with 906 tools, most prominently featuring game, finance, and science, among other domains.
Among the total of 1,749 examples, we have 575 \textit{perfect} examples,  599 with {\it under-specified} queries, and 575 examples with unavailable tools --- 261 of them are {\it human-replaceable}, and 314 are {\it non-replaceable}. See \S\ref{app:tool-dist} for detailed distribution of human- and non-replaceable tools in individual categories.

\begin{figure}[h!] 
% \vspace{-2mm}
    \centering
    \includegraphics[width=0.47\textwidth]{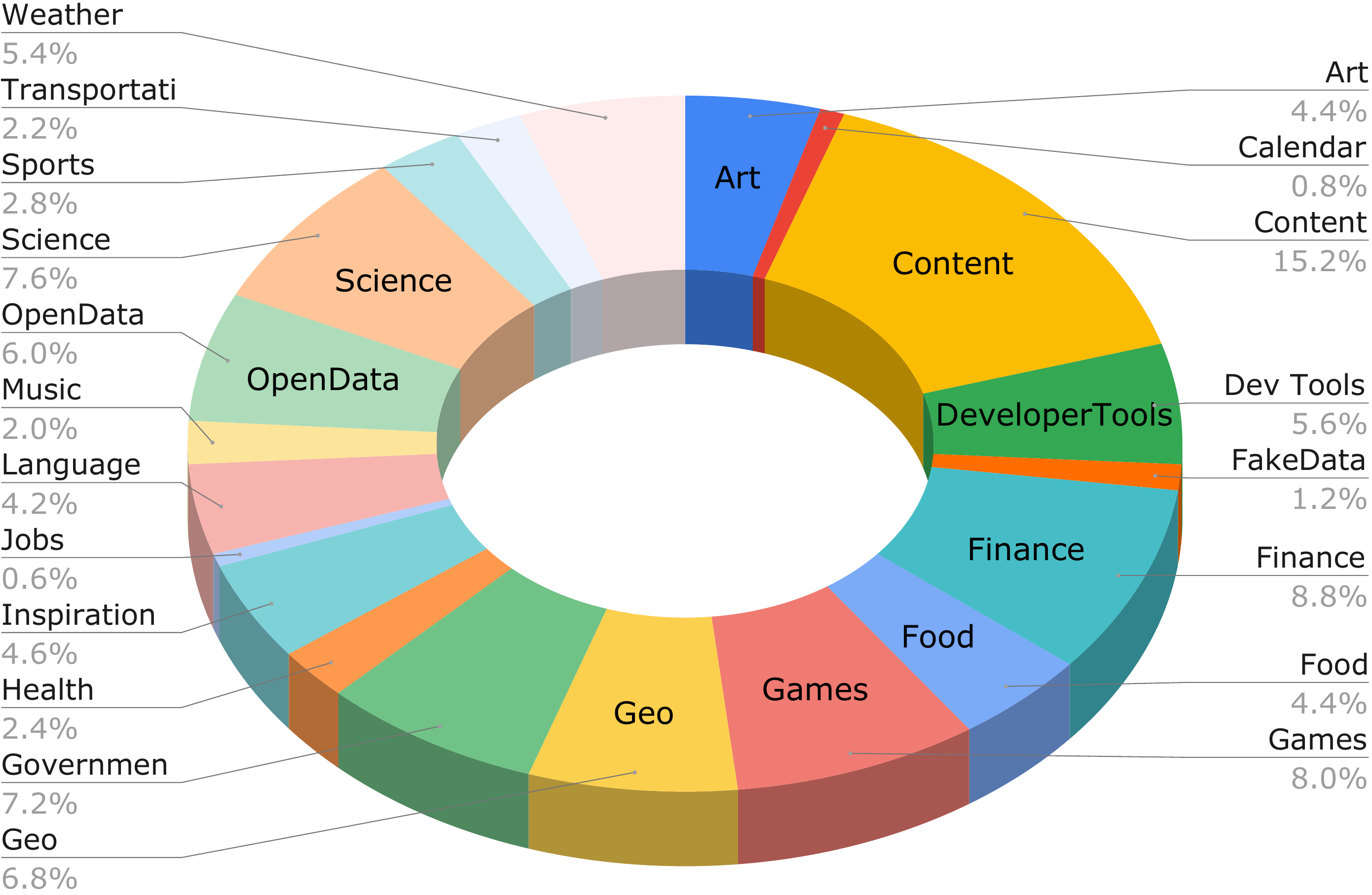}
    \vspace{-1mm}
    \caption{Category statistics of \ds queries.}
    \label{fig:data-dist}
    \vspace{-2mm}
\end{figure}

%%%%%%%%%%%%%%%%%%%%%%%%%%%%%%%%%%%%%%%%%%%%%%%%%%%%%%%%%%%%
\subsection{Evaluation Metrics}
\label{sec:3.4:eval-metrics}

We introduce the evaluation metrics regarding task success, awareness of missing components, among other dimensions.

\subsubsection{Correctness: Pass Rate}
We adopt the pass rate metric from \citet{qin2023toolllm}, which calculates the proportion of successful tasks. 
Specifically, we pass in the initial user query, the tool outputs, and the final TaLM response to a GPT-4o model, and ask it to evaluate if the final response solves the user query. GPT-4o grades responses binarily as ``Pass'' (1) or ``Fail'' (0). The final score is calculated by a majority vote across 5 GPT-4o graders. Refer to \S\ref{app:b.1:pass-rate-eval} for the detailed prompts and \S\ref{app:c:human-agreement} for human validation of GPT evaluation quality.

\subsubsection{Awareness of Missing Components}
Besides task completion, we also assess if TaLMs can identify the missing information or tools.

\paragraph{Information Awareness}
For {\it under-specified queries}, we evaluate whether the TaLM can detect insufficient information. We prompt the TaLM to determine if the query provides enough information to complete the task by responding \texttt{yes}, \texttt{idk} (I don't know), or \texttt{no}. A successful identification of insufficient information occurs when the model answers either \texttt{idk} or \texttt{no}. Formally, we define:
\[
\text{Info Awareness} = \frac{\text{Number of \texttt{idk}'s and \texttt{no}'s}}{\text{Total number of examples}}
\]

\paragraph{Tool Awareness}
When required tools are unavailable or non-functioning,\footnote{Non-functioning tools are those that cannot fulfill the requested operation, either because they are not accessible or lack the functionality needed for the task at hand.} we measure the TaLM's ability to recognize this limitation. To evaluate this, we prompt the model with: ``Do you have the right tools to complete the task?'' and ask for a \texttt{yes}, \texttt{idk}, or \texttt{no} response. Similar to Information Awareness, successful identification of tool unavailability occurs when the model answers either \texttt{idk} or \texttt{no}. Unlike Information Awareness, however, this metric specifically targets the model's recognition of tool-based limitations rather than informational gaps. For the non-replaceable setting, we thus have:
\[
\text{Tool Awareness} = \frac{\text{Number of \texttt{idk}'s and \texttt{no}'s}}{\text{Total number of examples}}
\]

See \S\ref{app:b.2:tool-aware-eval} for detailed prompts and evaluations.

\subsubsection{Unexpected Success}

In addition to cases where LMs correctly identify missing components and fail gracefully, an interesting scenario is when TaLMs unexpectedly solve the task correctly, despite lacking certain required information or tools.

Across examples with under-specified queries and unavailable tools, TaLMs are expected to respond \texttt{idk} or \texttt{no} to awareness questions. However, if the TaLM responds \texttt{yes} (to either information or tool awareness) and achieves a \textit{pass rate} of 1, this outcome is noteworthy. We thus compute the unexpected success by:
\[
\text{Unexpected Success} = \frac{\text{ \texttt{yes}
and pass rate = 1}}{\text{Total number of examples}}
\]

\subsubsection{Skipped Queries}
A query is skipped if the model responds \texttt{no} to an awareness question. We explicitly prompt the TaLM to respond no if it confidently believes that it lacks the necessary tools or information to solve a given task, and inform it that this decision will skip the task. We calculate the ratio of skipped examples among all examples. A lower score indicates fewer skipped queries, and higher model confidence.

%%%%%%%%%%%%%%%%%%%%%%%%%%%%%%%%%%%%%%%%%%%%%%%%%%%%%%%%%%%%
\begin{table*}[h!]
\centering
\small
\resizebox{\textwidth}{!}{
\begin{tabular}{l|cccc|cccc|cccc}
\toprule
\multirow{2}{*}{\textbf{Setting}} & \multicolumn{4}{c|}{\textbf{Claude-3.5-sonnet}} & \multicolumn{4}{c|}{\textbf{GPT4o}} & \multicolumn{4}{c}{\textbf{Qwen-72B-Instruct}} \\
\cmidrule{2-13}
{} & \textit{PR} & \textit{Aware} & \textit{Unexp} & \textit{Skip} & \textit{PR} & \textit{Aware} & \textit{Unexp} & \textit{Skip} & \textit{PR} & \textit{Aware} & \textit{Unexp} & \textit{Skip} \\
\midrule
Perfect           & 0.67 & 0.94 & 0.00 & 0.00 & 0.68 & 1.00 & 0.00 & 0.00 & 0.54 & 0.94 & 0.00 & 0.00 \\
Under-specified   & 0.31 & 0.42 & 0.24 & 0.08 & 0.36 & 0.18 & 0.33 & 0.05 & 0.40 & 0.08 & 0.31 & 0.05 \\
Unavailable tools & 0.25 & 0.70 & 0.03 & - & 0.28 & 0.06 & 0.05 & - & 0.15 & 0.03 & 0.04 & - \\
$~~~$non-replaceable & 0.09 & 0.85 & 0.04 & 0.09 & 0.11 & 0.09 & 0.09 & 0.06 & 0.06 & 0.05 & 0.05 & 0.03 \\
$~~~$replaceable & 0.41 & 0.54 & 0.01 & - & 0.44 & 0.03 & 0.01 & - & 0.23 & 0.00 & 0.02 & - \\
No-tools          & 0.10 & - & 0.00 & 0.00 & 0.29 & - & 0.01 & 0.00 & 0.09 & - & 0.01 & 0.00 \\
\bottomrule
\end{tabular}
}
\vspace{2mm}
\resizebox{\textwidth}{!}{
\begin{tabular}{l|cccc|cccc|cccc}
\toprule
\multirow{2}{*}{\textbf{Setting}} & \multicolumn{4}{c}{\textbf{Llama 8B}} & \multicolumn{4}{c|}{\textbf{Llama 70B}} & \multicolumn{4}{c}{\textbf{Llama 405B}} \\
\cmidrule{2-13}
{} & \textit{PR} & \textit{Aware} & \textit{Unexp} & \textit{Skip} & \textit{PR} & \textit{Aware} & \textit{Unexp} & \textit{Skip} & \textit{PR} & \textit{Aware} & \textit{Unexp} & \textit{Skip} \\
\midrule
Perfect           & 0.28 & 1.00 & 0.00 & 0.01 & 0.31 & 0.99 & 0.01 & 0.01 & 0.53 & 1.00 & 0.00 & 0.00 \\
Under-specified   & 0.14 & 0.02 & 0.13 & 0.00 & 0.29 & 0.19 & 0.29 & 0.11 & 0.25 & 0.11 & 0.25 & 0.06 \\
Unavailable tools & 0.11 & 0.02 & 0.02 & - & 0.07 & 0.36 & 0.10 & - & 0.24 & 0.02 & 0.07 & - \\
$~~~$non-replaceable & 0.03 & 0.02 & 0.03 & 0.00 & 0.04 & 0.55 & 1.00 & 0.38 & 0.12 & 0.02 & 0.12 & 0.02 \\
$~~~$replaceable & 0.19 & 0.01 & 0.01 & - & 0.10 & 0.17 & 0.19 & - & 0.36 & 0.02 & 0.01 & - \\
No-tools          & 0.00 & - & 0.00 & 0.00 & 0.37 & - & 0.02 & 0.00 & 0.28 & - & 0.06 & 0.00 \\
\bottomrule
\end{tabular}
}
\vspace{-2mm}
\caption{Performance of Claude, GPT, Qwen, and Llama models (8B, 70B, 405B) on \ds under uncertain and partial information settings. \textit{PR} stands for pass rate, \textit{Aware} refers to information/tool awareness, \textit{Unexp} stands for unexpected outcomes, and \textit{Skip} refers to skipped queries. 
}
\label{tab:main-results}
\vspace{-2mm}
\end{table*}

\section{Experiments and Results}\label{sec:5:results}
In this section, we introduce the experimental setup (\S\ref{sec:5.1:exp-setup}) and TaLM pefromance without \qaq (\S\ref{sec:5.2:wo-qaq}) and with (\S\ref{sec:4:qaq-method}) human assistance via \qaq.

\subsection{Experimental Setup}
\label{sec:5.1:exp-setup}
We evaluated five models including the openweight Qwen-72B-Instruct and Llama models of various sizes (8B, 70B, and 405B); we also benchmark multiple closed API models: GPT-4o and Claude-3.5-Sonnet. We use the default temperature
of t = 1.0 and sample n = 1 responses. We evaluate models on each split in \ds with the metrics specified in \S\ref{sec:3.4:eval-metrics}.

We report all results on \ds in \autoref{tab:main-results}.

\subsection{Standalone TaLMs}
\label{sec:5.2:wo-qaq}

\paragraph{Standard `Perfect' Baseline}
In the standard `perfect' setting, all models exhibited high awareness from 94--100\%, correctly identifying that they had all the necessary information and tools to solve the tasks.
Despite this high self-awareness, pass rates are lower than perfect --- GPT-4o at 68\%, Claude at 67\%, Qwen at 54\%, Llama 405 at 53\%, Llama 70B at 31\%, and Llama 8B at 28\%; which correlate well with their general abilities in problem-solving.

\paragraph{Awareness of Missing Components}
Across settings with either under-specified queries or unavailable tools, Claude achieves substantially higher awareness of missing information --- an average of 42\%, which is 23--40\% higher than other models, in particular, even 24\% higher than the top-performing GPT-4o. 
Similarly, in both non-replaceable and replaceable tool settings, Claude significantly outperforms other models and achieves a 70\% awareness, which is 34-68\% higher than other models and 64\% higher than GPT-4o.

Among Llama models, the medium-sized 70B model exhibits the highest awareness in the under-specified query split of 19\%, which is 17\% and 8\% higher than its 8B and even 405B counterparts. 
This trend continues in the unavailable tool setting, where the 70B model attains 36\% awareness, and gets 34\% higher than both the 8B and 405B models. In comparison, the other open-source candidate, Qwen, obtains relatively low awareness around 3\% with unavailable tools, mirroring the difficulty in recognizing missing components in weaker Llama models.

\paragraph{Is Awareness Related to Pass Rate?}
In the
under-specified setting, all TaLMs except Claude,
predominantly responded yes during the awareness assessment 82\%–98\% of the time, as shown in  \ref{tab:main-results}.
This suggests the Llama, Qwen, and GPT-4o models often confidently proceed despite uncertainty being present. Therefore there is no clear correlation between awareness and pass rate, and more of a characteristic of individual models.

Similarly in the {\it unavailable tools} setting shown in \autoref{tab:main-results}, Claude identifies 64\% more of the missing tools than GPT-4o, yet scores 3\% lower in pass rate than GPT-4o, suggesting a better judgment in knowledge sufficiency yet less success in solving the task under sub-optimal settings. 

Comparing Llama models with varied sizes, increasing model size from 8B to 70B increases the awareness score by 34\%, yet further increasing model size to 405B drags down the awareness score by 8\%, indicating decreased abilities to identify insufficient information, and behavior proceeding with the task even with insufficient content. Despite this loss in awareness, the pass rate increases up to 24\% with the largest 405B model.

Overall, strong models such as GPT-4o and Llama-405B may overestimate their capabilities, engaging in tasks they cannot solve, while Claude and smaller Llama models often approach unsolvable tasks more cautiously.

\paragraph{Comparing Human Replaceable and Non-Replaceable Tools}
Compared to non-replaceable tools, all models provided with human-replaceable tools show decreases in tool awareness by 1--34\%. In the meantime, models present substantial increases in pass rate by 6--36\%. Despite insufficient information, the nature of human-replaceable tools deceives the model into thinking they have sufficient information and thus can more easily solve the task. We conjecture this is due to the easier functionalities human-replaceable tools have, e.g., \texttt{get\_weather(location)}, compared to complex tools such as \texttt{rocket\_simulator()} which humans cannot easily replace.

\paragraph{Unexpected Success When Queries are Under-Specified}
Unexpected success should be minimal in all settings. This hypothesis holds in examples with unavailable tools, where all models achieve less than $5$\% of the cases correct. 
However, when user queries are under-specified, we surprisingly observe that all models obtain a substantial amount of unexpected successes, most prominently with GPT-4o solving 33\% of the cases correctly. 

We investigate this intriguing phenomenon and found models often refer to normal pragmatics and accidentally correctly assume the missing information to solve the task.
For instance, given a user query ``How is the stock performance today?'' with the company name under-specified. However, the model may pragmatically assume the missing reference to be {S\&P 500 Index}, since it is the most widely followed U.S. market benchmark. Despite the ambiguity, the model pragmatically fills in the missing information and solves the task correctly yet unexpectedly.

\section{Human-Assisted TaLMs}
\label{sec:4:qaq-method}

%%%%%%%%%%%
\subsection{The Ask-and-Help (\qaq) Setting}
One approach to alleviate TaLM failures under issues presented in \ds is to obtain assistance from humans.
Therefore, to examine how much human intervention can alleviate TaLM failures, we also experiment in a setting where TaLMs can request human assistance as an interactive tool. We refer to this setup as Ask-and-Help (\qaq).

The TaLM can choose to invoke \qaq at any time during inference, by querying humans as calling a tool with a textual argument $a$, e.g., ``What is your name'' for the example in \autoref{fig:taLM_example} (top).
The human then responds to $a$ with additional information or solutions, e.g., ``Mike'', which would be returned to the TaLM, much like the response from a traditional tool call. The TaLM then generates its final response based on human-provided information and other tool-calling or intermediate reasoning outputs. Since the pass rate is calculated using the final response of the TaLM, and \qaq occurs within its normal tool calling process, our interpretation of our other metrics does not change from the basic setup.

\paragraph{Additional Evaluation: Interaction Ratio}
We also measure the number of examples where TaLM chooses to interact with humans via \qaq, showing its awareness and willingness to seek help.

%%%%%%%%%%%
\begin{table*}[h!]
\centering
\small
\resizebox{\textwidth}{!}{
\begin{tabular}{l|cccc|cccc|cccc}
\toprule
\multirow{2}{*}{\textbf{Setting}} & \multicolumn{4}{c|}{\textbf{Claude-3.5-sonnet}} & \multicolumn{4}{c|}{\textbf{GPT4o}} & \multicolumn{4}{c}{\textbf{Qwen-72B-Instruct}} \\
\cmidrule{2-13}
{} & \textit{PR} & \textit{Aware} & \textit{Unexp} & \textit{Inter} & \textit{PR} & \textit{Aware} & \textit{Unexp} & \textit{Inter} & \textit{PR} & \textit{Aware} & \textit{Unexp} & \textit{Inter} \\
\midrule
Perfect & 0.67 & 0.94  & 0.00 & -  & 0.68 & 1.00  & 0.00 & -  & 0.54 & 0.98  & 0.00 & - \\
\midrule
Under-specified   & 0.61 & 0.51  & 0.31 & 0.61 & 0.61 & 0.21  & 0.47 & 0.58 & 0.48 & 0.15 & 0.34 & 0.29 \\
\midrule
Unavailable tools & 0.25 & 0.68 & 0.03 & 0.23 & 0.28 & 0.04 & 0.06 & 0.12 & 0.15 & 0.03  & 0.05  & 0.14 \\
$~~$non-replaceable   & 0.07 & 0.85  & 0.03 & 0.27 & 0.10 & 0.05  & 0.10 & 0.09 & 0.05 & 0.03 & 0.05 & 0.15  \\
$~~$replaceable       & 0.43 & 0.51  & 0.03 & 0.19 & 0.45 & 0.02  & 0.02 & 0.15 & 0.29 & 0.00    & 0.05  & 0.13 \\
\midrule
No-tools          & 0.10 & -     & 0.00 & - & 0.029 & -     & 0.01 & - & 0.09 & -    & 0.01  & -    \\
\bottomrule
\end{tabular}
}
\vspace{2mm}
\resizebox{\textwidth}{!}{
\begin{tabular}{l|cccc|cccc|cccc}
\toprule
\multirow{2}{*}{\textbf{Setting}} & \multicolumn{4}{c}{\textbf{Llama 8B}} & \multicolumn{4}{c|}{\textbf{Llama 70B}} & \multicolumn{4}{c}{\textbf{Llama 405B}} \\
\cmidrule{2-13}
{} & \textit{PR} & \textit{Aware} & \textit{Unexp} & \textit{Inter} & \textit{PR} & \textit{Aware} & \textit{Unexp} & \textit{Inter} & \textit{PR} & \textit{Aware} & \textit{Unexp} & \textit{Inter} \\
\midrule
Perfect           & 0.28 & 1.00 & 0.00  & -    & 0.31 & 0.99  & 0.01  & -    & 0.53 & 1.00 & 0.00  & -    \\
\midrule
Under-specified   & 0.24 & 0.01   & 0.24 & 0.24   & 0.33 & 0.25  & 0.24 & 0.21   & 0.53 & 0.02   & 0.53 & 0.25   \\
\midrule
Unavailable tools & 0.08  & 0.02   & 0.02  & 0.23   & 0.05  & 0.13  & 0.05  & 0.21   & 0.20 & 0.05   & 0.09  & 0.20   \\
$~~$non-replaceable & 0.03  & 0.03   & 0.03  & 0.26   & 0.05  & 0.21  & 0.05  & 0.18   & 0.16 & 0.09   & 0.15 & 0.16   \\
$~~$replaceable   & 0.13 & 0.01   & 0.00  & 0.19   & 0.04  & 0.04   & 0.05  & 0.24   & 0.38 & 0.01   & 0.02  & 0.24   \\
\midrule
No-tools          & 0.00  & -   & 0.00  & 0.00    & 0.37 & -   & 0.00  & -    & 0.28 & -   & 0.06  & -    \\
\bottomrule
\end{tabular}
}
\vspace{-2mm}
\caption{Performance of Claude, GPT, Qwen, and Llama models (8B, 70B, 405B) on \ds under uncertain and partial information settings with \textbf{AAH} assistance. \textit{Inter} refers to interaction rate.}
\label{tab:combined-results}
\vspace{-2mm}
\end{table*}

%%%%%%%%%%%%%%%%%%%%%%%%%%%%
\subsection{TaLM Performance with \qaq}
\label{sec:5.3:with-qaq}

We explored the impact of offering TaLMs the ability to interact with a human via the \qaq method on \ds. Effective use of \qaq depends on the model's ability to recognize when assistance is needed and to interact appropriately.

\paragraph{Interaction Improves Under-Specified Tasks}
After being augmented with \qaq, the \textit{human-replaceable} setting is now equipped with sufficient information. Correspondingly, strong models, including GPT-4o, Claude, and Llama 405B show slight increases in PR with the human interaction by 1--2\% as in \autoref{tab:main-results} and \autoref{tab:combined-results}. However, smaller models do not observably improve with human assistance, even with the missing tools that can be easily replaced by humans.

In contrast, the performance of \qaq-assisted TaLM on the \textit{under-specified query} setting shows significant pass rate improvements, by 25\%, 30\%, and 28\% on GPT-4o, Claude, and Llama-405B.
For Llama models with varying sizes, we also observe larger improvements as the model size grows, increasing from 10\% at 8B to 28\% at 405B.
Notably, the pass rate of Llama-405B with under-specified queries matched its perfect setting--- a 53\% pass rate. Moreover, Llama-70B surpasses its perfect setting pass rate by 2\%. Similarly, Qwen also shows moderate gains 6--8\% following the trend observed in the smaller Llama models.

For awareness, all models except Llama 405B increase by 3--9\%, suggesting that enabling human interaction may affect models' self-uncertainty assessment, leading to increased awareness of missing information. 
Lastly, the unexpected success increases by 7--28\% across all TaLMs except Llama 70B, because without human assistance, success in this setting is otherwise unexpected.

\paragraph{Human Interaction versus Awareness}
All models generally interact with humans via \qaq. Particularly when given under-specified queries, GPT and Claude models show substantially higher interaction rates than other models or than on other data splits, suggesting that they are eager to utilize human assistance to resolve incomplete information, as seen in \autoref{tab:combined-results}.

Across all models and settings, we do not find clear associations between the awareness of missing components, versus the interaction rate to \qaq. Rather, interactivity and awareness are more dependent on the model itself.
In the \textit{non-replaceable setting}, Claude's awareness remained high at 85\%, but its interaction rate was only 27\%, indicating limited use of human assistance despite recognizing high uncertainty. 

Despite recognizing high uncertainty (i.e., an 85\% awareness), Claude does not proportionally increase its use of \qaq nor achieve a higher pass rate via interaction, suggesting misalignment between self-uncertainty and the decision to seek help.

\section{Related Works}

\paragraph{TaLM Benchmarks} 

Most benchmarks about tool-augmented LMs collect tools from existing platforms that allow offline \citep{yang2023gpt4toolsteachinglargelanguage,xu2023toolmanip} or online executions \citep{li2023apibank,chen2024teval}, yet may bring degradation issues, where many tools become outdated over time \citep{guo2024stable}. More recent datasets emphasize tool diversity and realistic use cases \citep{qin2023toolllm,patil2023gorilla,li2023apibank,tang2023tool}, yet still assume perfect information and tool availability.
In contrast, our work studies failures with under-specified queries and unavailable tools, and alleviates them via human interaction.

\paragraph{Tool Failures in Practice}
Many TaLM works assume a perfect tool execution environment and user query specification, which no longer holds when used in practice. On the one hand, NL queries are often under-specified \citep{min2020ambigqa}, necessitating models to ask more information to proceed with the task. On the other hand, tools can deprecate over time \citep{qin2023toolllm}, suddenly broken \citep{guo2024stable}, or even return unpredictably false outputs \citep{sun2024toolsfaildetectingsilent}. Some works propose to use LM as a neural simulator of tool execution \citep{kim2023llm,guo2024stable} to maintain the perfect tool-availability assumption. Our work, instead, directly reveals and tackles both practical issues and proposes \qaq as an attempted solution.

\paragraph{Human-Model Interactive Problem Solving}
Many TaLM are designed to operate autonomously \citep{wang2024trove} without seeking help from other sources such as human users. However, as the tasks become more complex and the environment runs more dynamically \citep{guo2024stable}, it sometimes becomes theoretically impossible for the TaLM to finish the task itself. Human-in-the-loop comes as a useful technique \citep{mosqueira2023human}, especially in risk-critical tasks where human feedback or supervision is required. We thus employ this human interaction as an exploratory approach to alleviate our identified TaLM failures.

\section{Conclusion}

In this paper, we introduced \ds, a comprehensive benchmark designed to evaluate tool-augmented language models (TaLMs) under realistic conditions where user queries are under-specified or necessary tools are unavailable, consisting of 1,749 queries using 906 authorization-free tools across 21 categories.
Our experiments with both proprietary models (GPT, Claude) and open-weights models (LLaMa series) revealed that most TaLMs struggle to recognize missing information or unavailable tools.
To address this, we examine the Ask-and-Help (\qaq) method, allowing TaLMs to interact with humans in real time to obtain missing information or substitute non-functional tools. While we find \qaq improves the pass rate on under-specified queries, it has minimal impact when complex tools are unavailable.

\section*{Limitations}
While \ds provides a substantial foundation for evaluating the practical failures of TaLMs, we primarily focus on two failure modes: under-specified queries and unavailable tools. Other potential risk issues, such as adversarial inputs, are not addressed and could be explored in future work. The \qaq method involves human interaction, which may not be scalable or practical in all deployment scenarios due to concerns about latency, cost, or privacy. Implementing such a system in real-world applications would require careful consideration of these factors.

\section*{Author Contributions}\label{contributions}
Eduardo Treviño \textsuperscript{*} developed the \textit{AAH} method, led codebase development, conducted experiments, ran analyses in results section, created evaluations metrics (i.e \textit{Awareness}). Hugo Contant \textsuperscript{*} API testing and experimentation lead to \textit{Tool Validation} framework, assisted in analysis, experiments, and articulated \textit{Non-replaceable} vs \textit{Human-replaceable} setting. James Ngai assisted with code development, ran experiments, performed manual evaluations, and reviewed the manuscript. Zora Zhiruo Wang and Graham Neubig provided experimental guidance and revised the manuscript.

\textbf{{*} Indicates Co-First Authors}

\bibliography{custom}

\appendix

\twocolumn
\section{Benchmark Construction}
\label{app:a:benchmark-construction}
In this section, we highlight exact details of the components required during the construction of the benchmark.

%%%%%%%%%%%%%%%%%%%%%%%%%%%%%%%%%%%%%%%%%%%
\subsection{Tool Environment Construction}
\label{app:a.1:tool-environment-prompt}
In this section, we describe the prompt we provide GPT-4o for constructing the tool environment. 
The instruction generates a tool request Python code, tool unit test cases, and a tool representation which includes metadata in JSON format based on a tools documentation. It iterates over all tool documentations files and, for each tool, it generates these three files.
All tools utilized are covered under the following licenses: Creative Commons, MIT License, GNU General Public License (GPL) 2.0 or later, Open Data Commons Open Database License, Database Contents License, openFDA, Apache 2.0 License, Massachusetts Department of Transportation Developers License Agreement, GNU GPLv3, and ISC License.

\noindent \textbf{System Prompt}
\begin{lstlisting}[style=plain]
You are a helpful assistant designed to generate Python code, test cases, and metadata JSON files based on a tool documentation. Your task is to create Python functions to interact with all relevant tool URL's that a human might need based on the tool's documentation. 
Ensure that the function names are properly formatted and include necessary parameters. Additionally, generate corresponding test cases to verify the tool's functionality, and create a JSON file with metadata about the tool.
\end{lstlisting}

\noindent \textbf{Prompt}
\begin{lstlisting}[style=plain]
The following is documentation for a tool called "{tool_name}". Your task is to create a Python file "tool.py" to make requests to all the relevant tools that a human needs the functionality for based on the tool's documentation provided. Note: the tools function names should be lowercase and never start with a number.
Please ensure there are defaults in place (especially IDs or resource tags, etc., that are specific to the tool's URL). Additionally, ensure you create Test Cases separately to verify the tool's URL's work "tool_test.py".
Finally, create a metadata JSON file that provides metadata about the tool and all of its available endpoints "{tool_name}.json".
Here is an example tool file for a tool named artchicago: {seed_api_example}
Here is an example corresponding JSON file. Note how the names of the tool's in the `tool_list` match the function names in the Python code calling the tool URL: {seed_json_example}
Now, please do this for the tool named "{tool_name}". To capture your output generation, be sure to bold the titles, i.e., ### api.py then ```python or ```json before the code block:
"""{documentation_content}"""
\end{lstlisting}

%%%%%%%%%%%%%%%%%%%%%%%%%%%%%%%%%%%%%%%%%%%
\subsection{Tool Request}
\label{app:a.2:tool-request}
The following is an example of a tool request python file.
\begin{lstlisting}[language=Python, basicstyle=\small\ttfamily]
import requests
from typing import Optional, List

BASE_URL = "https://api.irail.be"

def stations(format: str = "json", lang: str = "en",  ):
    """
    Retrieve a list of all stations.

    :param format: The response format (json, xml, jsonp).
    :param lang: The language of any text or names in the response.
    """
    url = f"{BASE_URL}/stations/"
    params = {
        'format': format,
        'lang': lang,
    }
    response = requests.get(url, params=params)
    try:
        return response.json()
    except Exception as e:
        return {"error": str(e), "response": response.text}


def liveboard(station: str, id: Optional[str] = None, date: Optional[str] = None, time: Optional[str] = None,
              arrdep: str = "departure", lang: str = "en", format: str = "json", alerts: bool = False,
               ):
    """
    Retrieve a liveboard for a specified station.

    :param station: The name of the station to query.
    :param id: Optional station ID.
    :param date: Date for query, formatted as ddmmyy.
    :param time: Time for query, formatted as hhmm.
    :param arrdep: Whether to retrieve departures or arrivals.
    :param lang: The language of the response.
    :param format: The output format (json, xml, jsonp).
    :param alerts: Whether to include alerts.
    """
    url = f"{BASE_URL}/liveboard/"
    params = {
        'station': station,
        'id': id,
        'date': date,
        'time': time,
        'arrdep': arrdep,
        'lang': lang,
        'format': format,
        'alerts': alerts
    }
    response = requests.get(url, params=params)
    try:
        return response.json()
    except Exception as e:
        return {"error": str(e), "response": response.text}


def connections(from_station: str, to_station: str, date: str, time: str, timesel: str = "departure",
                format: str = "json", lang: str = "en", typeOfTransport: str = "automatic", alerts: bool = False,
                results: int = 6,  ):
    """
    Get routes between two stations, including realtime data on delays.

    :param from_station: The departure station.
    :param to_station: The destination station.
    :param date: Date for the query, formatted as ddmmyy.
    :param time: Time for the query, formatted as hhmm.
    :param timesel: Whether results should show arrivals or departures.
    :param format: The response format.
    :param lang: The language of the response.
    :param typeOfTransport: Types of transport to include.
    :param alerts: Include alerts or not.
    :param results: Number of results to return.
    """
    url = f"{BASE_URL}/connections/"
    params = {
        'from': from_station,
        'to': to_station,
        'date': date,
        'time': time,
        'timesel': timesel,
        'format': format,
        'lang': lang,
        'typeOfTransport': typeOfTransport,
        'alerts': alerts,
        'results': results
    }
    response = requests.get(url, params=params)
    try:
        return response.json()
    except Exception as e:
        return {"error": str(e), "response": response.text}


def vehicle(id: str, date: Optional[str] = None, format: str = "json", lang: str = "en", alerts: bool = False,
             ):
    """
    Retrieve information about a vehicle including stops and delays.

    :param id: The ID of the vehicle.
    :param date: Date for the query, formatted as ddmmyy.
    :param format: The response format.
    :param lang: The language of the response.
    :param alerts: Include alerts or not.
    """
    url = f"{BASE_URL}/vehicle/"
    params = {
        'id': id,
        'date': date,
        'format': format,
        'lang': lang,
        'alerts': alerts
    }
    response = requests.get(url, params=params)
    try:
        return response.json()
    except Exception as e:
        return {"error": str(e), "response": response.text}


def composition(id: str, format: str = "json", data: str = "", lang: str = "en",
                 ):
    """
    Retrieve the composition of a train, i.e., carriages and locomotives.

    :param id: The ID of the train.
    :param format: The response format.
    :param data: To get all raw unfiltered data use 'all'.
    :param lang: The language of the response.
    """
    url = f"{BASE_URL}/composition/"
    params = {
        'id': id,
        'format': format,
        'data': data,
        'lang': lang
    }
    response = requests.get(url, params=params)
    try:
        return response.json()
    except Exception as e:
        return {"error": str(e), "response": response.text}


def disturbances(format: str = "json", lineBreakCharacter: str = "", lang: str = "en",
                  ):
    """
    Retrieve information about current disturbances.

    :param format: The response format.
    :param lineBreakCharacter: Character for line breaks in text.
    :param lang: The language of the response.
    """
    url = f"{BASE_URL}/disturbances/"
    params = {
        'format': format,
        'lineBreakCharacter': lineBreakCharacter,
        'lang': lang
    }
    response = requests.get(url, params=params)
    try:
        return response.json()
    except Exception as e:
        return {"error": str(e), "response": response.text}
\end{lstlisting}

%%%%%%%%%%%%%%%%%%%%%%%%%%%%%%%%%%%%%%%%%%%
\subsection{Tool Representation}
\label{app:a.3:tool-representation}
\begin{lstlisting}[language=Python, basicstyle=\small\ttfamily]
{
    "tool_name": "irail",
    "tool_description": "Tool to access railway time schedules in Belgium, including stations, liveboards, connections, vehicles, disturbances, and more.",
    "title": "iRail API",
    "pricing": "FREE",
    "score": {
        "avgServiceLevel": 95,
        "avgLatency": 150,
        "avgSuccessRate": 98,
        "popularityScore": 9.0,
        "__typename": "Score"
    },
    "home_url": "https://api.irail.be",
    "host": "api.irail.be",
    "api_list": [
        {
            "name": "stations",
            "url": "https://api.irail.be/stations/",
            "description": "Retrieve a list of all stations.",
            "method": "GET",
            "required_parameters": [],
            "optional_parameters": [
                {"name": "format", "type": "STRING", "description": "Response format", "default": "json"},
                {"name": "lang", "type": "STRING", "description": "Language of response", "default": "en"}
            ],
            "statuscode": 200
        },
        {
            "name": "liveboard",
            "url": "https://api.irail.be/liveboard/",
            "description": "Retrieve liveboard for a station including arrivals and departures.",
            "method": "GET",
            "required_parameters": [
                {"name": "station", "type": "STRING", "description": "Station name"}
            ],
            "optional_parameters": [
                {"name": "id", "type": "STRING", "description": "Station ID"},
                {"name": "date", "type": "STRING", "description": "Date for query"},
                {"name": "time", "type": "STRING", "description": "Time for query"},
                {"name": "arrdep", "type": "STRING", "description": "Arrivals or departures", "default": "departure"},
                {"name": "lang", "type": "STRING", "description": "Language of response", "default": "en"},
                {"name": "format", "type": "STRING", "description": "Response format", "default": "json"},
                {"name": "alerts", "type": "BOOLEAN", "description": "Include alerts", "default": "false"}
            ],
            "statuscode": 200
        },
        {
            "name": "connections",
            "url": "https://api.irail.be/c"
            "description": "Get routes between two stations.",
            "method": "GET",
            "required_parameters": [
                {"name": "from", "type": "STRING", "description": "Departure station"},
                {"name": "to", "type": "STRING", "description": "Destination station"}
            ],
            "optional_parameters": [
                {"name": "date", "type": "STRING", "description": "Date for query"},
                {"name": "time", "type": "STRING", "description": "Time for query"},
                {"name": "timesel", "type": "STRING", "description": "Arrivals or departures", "default": "departure"},
                {"name": "lang", "type": "STRING", "description": "Language of response", "default": "en"},
                {"name": "format", "type": "STRING", "description": "Response format", "default": "json"},
                {"name": "typeOfTransport", "type": "STRING", "description": "Type of transport", "default": "automatic"},
                {"name": "alerts", "type": "BOOLEAN", "description": "Include alerts", "default": "false"},
                {"name": "results", "type": "INTEGER", "description": "Number of results", "default": 6}
            ],
            "statuscode": 200
        },
        {
            "name": "vehicle",
            "url": "https://api.irail.be/vehicle/",
            "description": "Retrieve information about a vehicle.",
            "method": "GET",
            "required_parameters": [
                {"name": "id", "type": "STRING", "description": "Vehicle ID"}
            ],
            "optional_parameters": [
                {"name": "date", "type": "STRING", "description": "Date for query"},
                {"name": "lang", "type": "STRING", "description": "Language of response", "default": "en"},
                {"name": "format", "type": "STRING", "description": "Response format", "default": "json"},
                {"name": "alerts", "type": "BOOLEAN", "description": "Include alerts", "default": "false"}
            ],
            "statuscode": 200
        },
        {
            "name": "composition",
            "url": "https://api.irail.be/composition/",
            "description": "Retrieve the composition of a train.",
            "method": "GET",
            "required_parameters": [
                {"name": "id", "type": "STRING", "description": "Train ID"}
            ],
            "optional_parameters": [
                {"name": "format", "type": "STRING", "description": "Response format", "default": "json"},
                {"name": "data", "type": "STRING", "description": "Raw or filtered data", "default": ""},
                {"name": "lang", "type": "STRING", "description": "Language of response", "default": "en"}
            ],
            "statuscode": 200
        },
        {
            "name": "disturbances",
            "url": "https://api.irail.be/disturbances/",
            "description": "Retrieve information about current disturbances.",
            "method": "GET",
            "required_parameters": [],
            "optional_parameters": [
                {"name": "format", "type": "STRING", "description": "Response format", "default": "json"},
                {"name": "lineBreakCharacter", "type": "STRING", "description": "Line break character", "default": ""},
                {"name": "lang", "type": "STRING", "description": "Language of response", "default": "en"}
            ],
            "statuscode": 200
        }
    ]
}
\end{lstlisting}

%%%%%%%%%%%%%%%%%%%%%%%%%%%%%%%%%%%%%%%%%%%
\subsection{Python Unit Test Example}
\label{app:a.4:unittest}
\begin{lstlisting}[language=Python, basicstyle=\small\ttfamily]
import unittest
from api import stations, liveboard, connections, vehicle, composition, disturbances

class TestIRailAPI(unittest.TestCase):
    
    def test_stations(self):
        response = stations()
        self.assertIn('station', response)
    
    def test_liveboard(self):
        response = liveboard('Gent-Sint-Pieters')
        self.assertIn('station', response)
            
    def test_connections(self):
        response = connections('Gent-Sint-Pieters', 'Mechelen', '23082024', '1130')
        self.assertIn('connection', response)

    def test_vehicle(self):
        response = vehicle('BE.NMBS.IC3033')
        self.assertIn('vehicle', response)
    
    def test_composition(self):
        response = composition('S51507')
        self.assertIn('composition', response)
    
    def test_disturbances(self):
        response = disturbances()
        self.assertIn('disturbance', response)

if __name__ == '__main__':
    unittest.main()
\end{lstlisting}

\subsection{Query-tool Example}
\label{app:a.5:perfect-query-example}

Below is an example of a query-tool interaction involving tools from the Met Museum and the Art Institute of Chicago.

\begin{lstlisting}[language=python, basicstyle=\small\ttfamily]
[
  {
    "tool_list": [
        {
            "category_name": "Art",
            "tool_name": "metmuseum",
            "function_name": "search_objects",
            "tool_description": "Search for objects in the Met's collection",
            "required_parameters": [
                {
                    "name": "q",
                    "type": "STRING",
                    "description": "Search term",
                    "default": "Impressionist paintings"
                }
            ],
            "optional_parameters": [
                {
                    "name": "departmentId",
                    "type": "INTEGER",
                    "description": "ID of the department",
                    "default": "11"
                }
            ],
            "method": "GET",
            "template_response": {
                "total": "int",
                "objectIDs": ["int"]
            }
        },
      {
            "category_name": "Art",
            "tool_name": "artchicago",
            "function_name": "artworks_search",
            "tool_description": "Search artworks in the Art Institute of Chicago data in the aggregator. Artworks in the groups of essentials are boosted so they'll show up higher in results.",
            "required_parameters": [
                {
                    "name": "q",
                    "type": "STRING",
                    "description": "Your search query.",
                    "default": "monet"
                }
            ],
            "optional_parameters": [
                {
                    "name": "size",
                    "type": "INTEGER",
                    "description": "Number of results to return. Pagination via Elasticsearch conventions.",
                    "default": "10"
                },
                {
                    "name": "sort",
                    "type": "STRING",
                    "description": "Used in conjunction with query to sort results.",
                    "default": ""
                }
            ],
            "method": "GET",
            "template_response": {
                "pagination": {
                    "total": "int",
                    "limit": "int",
                    "offset": "int",
                    "total_pages": "int",
                    "current_page": "int"
                },
                "data": [
                    {
                        "id": "int",
                        "title": "str",
                        "artist_display": "str",
                        "place_of_origin": "str",
                        "date_display": "str",
                        "medium_display": "str",
                        "dimensions": "str",
                        "thumbnail": {
                            "alt_text": "str",
                            "width": "int",
                            "height": "int",
                            "iiif_url": "str"
                        }
                    }
                ]
            }
        }
    ],
    "query": "I want to find Impressionist paintings in the European Paintings department in the Met's collection. Additionally, can you find artworks related to Monet in the Art Institute of Chicago?",
    "query_id": 2
  }
]
\end{lstlisting}

%%%%%%%%%%%%%%%%%%%%%%%%%%%%%%%%%%%%%%%%%%%
\subsection{Perfect Queries Prompt}
\label{app:a.6:perfect-query-prompt}

Perfect queries are constructed by combining different components in the Tool Environment. The code processes the Tool Requests, which are Python files that handle interactions with tools, and Tool Representations, which contain important information about the tools, such as required parameters, optional parameters, and expected responses. The code systematically pairs tools from the same category and generates. We prompt \texttt{gpt-4o-2024-08-06} to generate a query combining these tools.

\noindent \textbf{Prompt}
\begin{lstlisting}[style=plain]
Below I have attached 2 Tools "{tool1}" and "{tool2}", which are Python files that make requests to the tools from the "{category_folder}" category, and their corresponding metadata JSON files that provide additional information about the tools, as well as unit tests that have been run on these tools. Utilize the parameters used in these unit tests, and the information about the tools to help you with your task. Your task is to create a tool-question JSON file that asks a question a human would ask. Note: For the tool-question JSON file, be sure to include the name of the tool function from the Python files inside the{tool_list}; they should be the same name and format as the function provided in the Python code.
"{tool_1}" tool: {tool_1_python_request}
"{tool_1}" unittest: {tool_1_python_unittest}
"{tool_1}" tool metadata: {tool_1_json_representation}

"{tool_2}" tool: {tool_2_python_request}
"{tool_2}" unittest:{tool_2_python_unittest}
"{tool_2}" tool metadata: {tool_2_json_representation}
tool-question JSON example: {query-tool_example}
\end{lstlisting}

\section{Statistical Details of \ds}
\label{app:tool-dist}

We study the distinction of human-replaceable and non-replaceable tools in \S\ref{sec:3.3:ds-overview}. Here we present the detailed distribution of these two types of tools across the 21 categories involved in our \ds.

\begin{figure*}[h!] 
    \centering
    \includegraphics[width=0.80\textwidth]{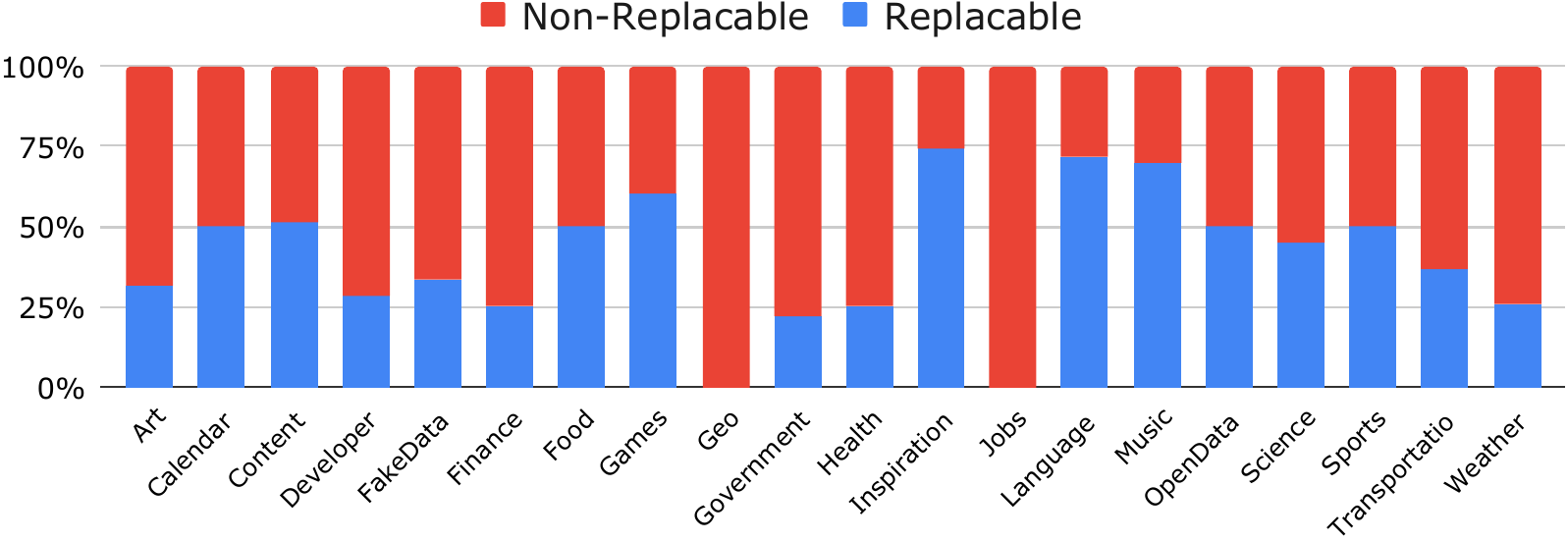}
    \caption{Distribution of human-replaceable and non-replaceable tools in each category in \ds.}
    \label{fig:data-dist}
\end{figure*}
\section{Model-based Evaluation}
\label{app:b:model-based-eval}

\subsection{Pass Rate}
\label{app:b.1:pass-rate-eval}

The pass rate metric evaluates whether the assistant's response successfully fulfills the user's instruction. The evaluation is performed by a grader model that assesses the assistant's reply and determines a pass or fail outcome. To ensure reliability, the evaluation is conducted multiple times (up to 5 attempts), and a majority voting mechanism is used to decide the final result. If the majority of evaluations result in a pass, the assistant's response is considered a pass; otherwise, it's a fail.

\textbf{Prompt:}

\begin{lstlisting}[style=plain]
System Message:
You are a grader. You will be given a user's instruction, assistant's answer, and the API's accessed along with their responses. Based on the answer given, determine if the instruction was completed or not. Always begin your response with either 'Pass' or 'Fail'.

Prompt:
Below I have attached a user's instruction, assistant's answer, and attached API's accessed with their responses. Did the assistant's answer complete the instruction given? Begin your response with either 'Pass' or 'Fail'.

Query: {query_text}
Answer: {assistant_reply}

API's accessed and responses:
{function_context}
\end{lstlisting}

\subsection{Information Awareness Prompt}
\label{app:b.2:info-aware-eval}
The system is tasked with determining whether it can gather, infer, or has enough information to answer the user’s query based on the following factors:
\begin{enumerate}
    \item Existing knowledge of the world, independent of the tools.
    \item The functionality of the available tools that the system has access to.
    \item A combination of both existing knowledge and tool functionality.
\end{enumerate}

When evaluating the query, the system considers whether:
\begin{itemize}
    \item The query provides enough information for a direct answer.
    \item The available tools can assist in obtaining or clarifying the necessary information.
    \item A combination of the tool's functionality and existing knowledge can help infer, gather, or produce the necessary information.
\end{itemize}

Based on this evaluation, the system responds as follows:
\begin{itemize}
    \item \textbf{Yes}: The system has all the information needed or can gather or infer it using either existing knowledge, available tools, or both. In this case, the system will attempt to answer.
    \item \textbf{IDK}: The system is uncertain but willing to attempt an answer using its knowledge, tools, or a combination of both.
    \item \textbf{No}: The system cannot answer the query with the available knowledge, tools, or any combination of both, and it will skip answering.
\end{itemize}

In all responses, the system must begin with \textbf{Yes}, \textbf{IDK}, or \textbf{No}, followed by a brief explanation of the reasoning. If the response is \textbf{No}, the system skips the query. If the response is \textbf{Yes} or \textbf{IDK}, the system will attempt to provide an answer.

\textbf{Prompt:}
\begin{lstlisting}[style=plain]
System message:
Your task is to determine if you can gather, infer, or have all the information needed to answer the user's query using:
1. Your existing knowledge (regardless of the tools).
2. The functionality of the available tools you have access to.
3. A combination of both.

When evaluating the query, consider whether:
The query provides enough information for you to answer directly.
The available tools can help you obtain the necessary information or clarify the query.
Or if a combination of a tool's functionality and your existing knowledge helps you infer, gather, or have the necessary information you need to answer.

Based on this evaluation:
Respond with 'Yes' if you can gather, infer, or have all the information needed to answer the query using your knowledge, the tools, or both.
Respond with 'IDK' if you are unsure but are willing to attempt an answer.
Respond with 'No' if you cannot answer the query at all, even with your knowledge and the tools available. This means the query is completely unanswerable given the current resources.
Important: Always start your response with 'Yes,' 'No,' or 'IDK,' followed by a brief explanation of your reasoning. If you respond with 'No,' you will skip attempting to answer the query. If you respond with 'Yes' or 'IDK,' you are willing to try to answer.

Prompt:
Based on the user's query, your knowledge of the world, and the functionality of the available tools, determine if you can gather, infer, or have all the information needed to answer the request. Remember: Start with 'Yes,' 'No,' or 'IDK,' followed by an explanation. 'Yes' means you have enough information, you can infer it, or can obtain it using the tools, and you will attempt to answer. 'IDK' means you are uncertain but willing to try using your knowledge, tools, or a combination of both. 'No' means you cannot answer the query with your knowledge, the available tools, or a combination of both and you will skip it. Query: {query}, Your knowledge of the world, Available Tools and Their Functionalities: {functions}
\end{lstlisting}

\paragraph{Tool Awareness Prompt}
\label{app:b.2:tool-aware-eval}
The system is tasked with determining whether it has the necessary tools, knowledge, or a combination of both to answer the user's query. The evaluation is based on the following considerations:
\begin{enumerate}
    \item The system’s existing knowledge, independent of the available tools.
    \item The functionality of the tools accessible to the system.
    \item A combination of both the existing knowledge and tool functionality.
\end{enumerate}

When evaluating the query, the system considers whether:
\begin{itemize}
    \item The functionality of the tools can help obtain the necessary answer.
    \item The system's existing knowledge allows for directly answering the query.
    \item A combination of tool functionality and existing knowledge is sufficient to provide an answer.
\end{itemize}

Based on this evaluation, the system will respond as follows:
\begin{itemize}
    \item \textbf{Yes}: The system has sufficient knowledge, tools, or a combination of both, and will attempt to answer.
    \item \textbf{IDK}: The system is uncertain but willing to try answering using its knowledge, tools, or both.
    \item \textbf{No}: The system cannot answer the query, even with the available knowledge, tools, or a combination of both, and will skip attempting to answer.
\end{itemize}

In all cases, the system must begin its response with \textbf{Yes}, \textbf{IDK}, or \textbf{No}, followed by a brief explanation. If the response is \textbf{No}, the system will skip the query. For \textbf{Yes} or \textbf{IDK}, the system will proceed to answer the query.

\textbf{Prompt:}
\begin{lstlisting}[style=plain]
System Message:
Your task is to determine whether you can answer the user's query using:

1. Your existing knowledge (regardless of the tools).
2. The functionality of the available tools you have access to.
3. A combination of both.

When evaluating the query, consider whether:
The functionality of the tools can help you obtain the answer.
Your existing knowledge allows you to answer directly.
Or if a combination of a tool's functionality and your existing knowledge helps you obtain the answer.

Based on this evaluation:
Respond with 'Yes' if you can answer the query using your knowledge, the tools, or a combination of both.
Respond with 'IDK' if you are unsure but are willing to attempt an answer.
Respond with 'No' if you cannot answer the query at all, even with your knowledge and the tools available, or a combination of both. This means the query is completely unanswerable given the current resources, and you will skip attempting to answer the query.

Important: Always start your response with 'Yes,' 'No,' or 'IDK,' followed by a brief explanation of your reasoning. If you respond with 'No,' you will skip attempting to answer the query. If you respond with 'Yes' or 'IDK,' you are willing to try to answer.

Prompt:
Based on the available tools' functionality and your knowledge of the world, determine whether you have the necessary tools, knowledge, or a combination of both to answer the query. Start with 'Yes,' 'No,' or 'IDK,' followed by an explanation. 'Yes' means you have the knowledge, tools, or a combination of both, and you will attempt to answer. 'IDK' means you are uncertain but willing to try to answer. 'No' means you cannot answer the query with your knowledge, the available tools, or a combination of both, and you will skip it. Query: {query}, Your knowledge of the world, Available Tools: {functions}
\end{lstlisting}

\section{Human Agreement with Model-Based Evaluation}
\label{app:c:human-agreement}
In \S\ref{sec:3.4:eval-metrics} we propose to use model-based evaluation to examine the correctness of model-generated responses, i.e., \textit{pass rate}. To validate the reliability of GPT-4o's pass rate evaluation, we analyzed 95 examples randomly sampled from GPT-4o’s responses on the perfect setting and compare the pass rate results between human evaluators and GPT-4o. The comparison showed an 83.2\% agreement in the correctness of model-generated final responses between GPT-4o and human evaluators. This high level of agreement indicates that GPT-4o provides evaluation results similar to those of humans, making it a credible evaluator capable of simulating human assessment of pass rates.
\section{Confidence Estimation of Results}
\label{app:conf-est}

% Our primary objective was to create a benchmark that reflects diverse real-world scenarios by covering a broad range of tools and tool categories. This approach necessitated a balance between breadth (number of tools and categories) and depth (number of queries per tool). While this resulted in an average of approximately two queries per tool and some categories with fewer than 25 samples, we believe that the diversity and coverage of our dataset enhance its overall representativeness and utility.

While our benchmark contains a total of 1,749 examples, we conduct power analysis and result confidence estimations to further verify its effectiveness in supporting the conclusions drawn in our experiment sections.

First, we conducted significance testing and power analysis to determine the minimum number of examples required to detect meaningful differences between our experimented models with adequate statistical power. We found that at most 92 queries are sufficient to achieve reliable detection of meaningful differences between configurations with medium effect size ($w=0.5$), a significance level of $\alpha=0.005$, and a statistical power of $0.90$. Our dataset contains 261–-599 examples in each data split, which is sufficiently higher than 92, suggesting that it can guarantee reliable and significant results.

% Regarding categories with fewer samples (e.g., Calendar, FakeData), we acknowledge that certain categories inherently have a smaller number of tools and thus fewer queries. These categories are included to ensure comprehensive coverage of various tool functionalities, even if they are less represented in terms of quantity. While statistical power in these smaller categories may be lower, they contribute to the overall diversity and comprehensiveness of the benchmark.

Moreover, to enhance the transparency and robustness of our evaluation results, we have included confidence intervals for our key metrics (pass rate, awareness, unexpected outcomes, and skipped queries) across all settings and models. Specifically, we calculate the 95\% confidence intervals using the Wilson score interval for proportions and report them in \autoref{tab:result-confidence}.

\begin{table*}[h!]
\centering
\small
\resizebox{0.93\textwidth}{!}{
\begin{tabular}{lcc|ccc}
\toprule
{\bf Metric} & {\bf Model} & {\bf Setting} & {Value (\%)} & {N} & {Confidence Interval (\%)} \\
\midrule
\multirow{3}{*}{Pass Rate} & {GPT-4o} & {under-specified} & {36.0} & {599} & {$[32.2, 39.9]$} \\
{} & {Claude-3.5-sonnet} & {perfect} & {67.0} & {575} & {$[63.1, 70.8]$} \\
{} & {LLaMa-70B} & {unavailable tools (non-replaceable)} & {12.0} & {314} & {$[~~8.6, 16.2]$} \\
\midrule
\multirow{3}{*}{Awareness} & {GPT-4o} & {under-specified} & {18.0} & {599} & {$[15.0, 21.3]$} \\
{} & {Claude-3.5-sonnet} & {under-specified} & {42.0} & {599} & {$[38.1, 46.0]$} \\
{} & {LLaMa-70B} & {unavailable tools} & {36.0} & {575} & {$[32.1, 40.1]$} \\
\bottomrule
\end{tabular}
}
\vspace{-2mm}
\caption{Confidence intervals of model performance.}
\label{tab:result-confidence}
\vspace{-2mm}
\end{table*}

\end{document}